\documentclass[b5paper,twoside]{jpconf}  
\usepackage{graphicx}

\begin{document}

\title[Magnetic Field Diffusion]{Magnetic Field Diffusion and the Formation
of Circumstellar Disks}  

\author[Basu et al.]{Shantanu Basu$^1$, Wolf B. Dapp$^{1,2}$, and Matthew W. Kunz$^3$}

\address{$^1$Department of Physics and Astronomy, University of Western Ontario, London, Ontario, N6A 3K7, Canada}
\address{$^2$J\"ulich Supercomputing Centre, Institute for Advanced Simulation, FZ J\"ulich, Germany}
\address{$^3$Department of Astrophysical Sciences, Princeton University, Princeton, NJ 08544, USA; Einstein Postdoctoral Fellow}

\ead{basu@uwo.ca} 

\begin{abstract}
A non-ideal MHD collapse calculation employing the axisymmetric thin-disk
approximation is used to resolve cloud core collapse down to the scales
of the second (stellar) core. Rotation and a magnetic braking torque are 
included in the model, and the partial ionization resulting in ambipolar diffusion and Ohmic dissipation is calculated from a detailed chemical 
network. We find that a centrifugal
disk can indeed form in the earliest stage of star
formation, due to a shut-off of magnetic braking caused by magnetic field 
diffusion in the first core region. Thus, there is no catastrophic
magnetic braking in a model with realistic non-ideal MHD.
\end{abstract}

\section{Results and Model Description}

Fig. 1 shows the magnetic field lines within the inner 10 AU at the end of the simulation in two
different cases. A second (stellar) core has just formed in both cases, 
after second collapse of the first hydrostatic core. The second core is
centered at the origin ($r=0,z=0$) and the thin-disk model extends in the
equatorial plane. The magnetic field lines are obtained above and below the
midplane using the force-free and current-free approximation, using the
currents in the midplane as a source term. 

The flux-frozen model (dashed lines) shows extreme pinching of the magnetic
field lines, into nearly a split-monopole configuration. This is due to the
dragging-in of field lines in the flux-freezing limit. The extreme flaring
of magnetic field lines results in catastrophic magnetic braking
in this model. Magnetic field lines that tie small radii near the protostar
with a much larger lever arm far above the disk can very efficiently 
extract angular momentum from the midplane region. No centrifugally-supported 
disk is able to form in this case, and radial
infall continues onto the protostar. Conversely, the non-ideal MHD model
(solid lines) shows that the same field lines straighten out significantly on 
the small scales shown in the figure (the full model extends to about 
10$^4$ AU in radius). The straightening of field lines on small scales 
shuts down the efficiency of magnetic braking, and a disk of radius
$\approx 10\, R_{\odot}$ is formed when the protostar has mass
$\approx 10^{-3}\,M_{\odot}$ and has just begun the accretion process. 
Based on the angular momentum content of the core at this time,
we estimate that direct infall will lead to a small disk (radius $< 10$ AU)
for $\approx 4 \times 10^4$ yr, representing the early Class 0 phase. 

This result extends previous work on disk formation that employed a 
simplified parametrization of just Ohmic dissipation \cite{dap10}. Here, 
we use a full chemical network model to obtain partial ionization
at each location in the cloud and thereby calculate local coefficients
of both ambipolar diffusion and Ohmic dissipation. We adopt the model of
\cite{kun09} in this regard. A detailed presentation of our results 
can be found in \cite{dap12}. 

\begin{figure}
\begin{center}
\includegraphics[width=3.5in]{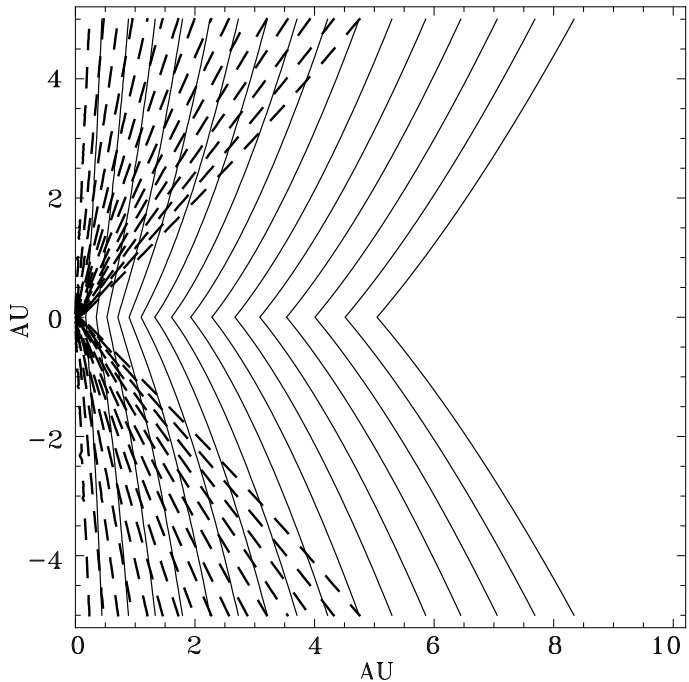}
\end{center}
\caption{Magnetic field lines. The box has dimensions of 10 AU on each side,
and the newly formed second core is located at the origin. The dashed lines 
represent a flux-frozen model, while the solid lines show the same 
field lines for a model including non-ideal MHD effects.
Note the extreme flaring of the magnetic field in the flux-frozen model
(leading to catastrophic magnetic braking) and the straightening out
of field lines in the non-ideal MHD model.}
\end{figure}


\section*{References}
\begin{thebibliography}{3} 
\bibitem{dap10} Dapp, W. B., \& Basu, S. 2010, \aap, 521, L56
\bibitem{kun09} Kunz, M. W., \& Mouschovias, T. Ch. 2009, \apj, 693, 1895
\bibitem{dap12} Dapp, W. B., Basu, S., \& Kunz, M. W. 2012, arXiv:1112.3801
\end{thebibliography}
\end{document}